\documentclass[]{aa}  
\usepackage{graphicx}
\graphicspath{{mes_figures/}}
\usepackage{natbib}
\usepackage{txfonts}
\usepackage{multirow}
\usepackage{slashbox}
\def\figwidth{\linewidth}

\begin{document}
   \title{Thermal phase curves of nontransiting terrestrial exoplanets}
   \subtitle{2. Characterizing airless planets}

   \author{A.S. Maurin
          \inst{1,2},
          F. Selsis\inst{1,2},
          F. Hersant\inst{1,2}, and
          A. Belu\inst{1,2}
          }
          
   \authorrunning{A.S. Maurin et al.}
 \titlerunning{Phase curves of non-transiting terrestrial exoplanets. 2. Airless planets}

   \institute{Universit\'e de Bordeaux, Observatoire Aquitain des Sciences de l'Univers,
2 rue de l'Observatoire, BP 89, F-33271 Floirac Cedex, France
\\
              \email{maurin@obs.u-bordeaux1.fr}
         \and
            CNRS, UMR 5804, Laboratoire d'Astrophysique de Bordeaux,
2 rue de l'Observatoire, BP 89, F-33271 Floirac Cedex, France \\
             }

   \date{Received 8 April 2011 / Accepted 13 October 2011}

 
  \abstract
  {The photometric signal we receive from a star hosting a planet is modulated by the variation in the planet signal with its orbital phase. Such phase variations (or phase curves) are observed for transiting hot Jupiters with current instrumentation and have also been measured for one transiting terrestrial planet (Kepler 10 b) and one nontransiting gas giant (Ups A b). Future telescopes (JWST and EChO) will have the capability of measuring thermal phase curves of exoplanets, including hot rocky planets in transiting and nontransiting configurations and at different wavelengths. Short-period planets with a mass below 10~R$_{\oplus}$ are indeed frequent, and nearby targets (within 10~pc) are already known and more are to be found. }
  {We test the possibility of using multiwavelength infrared phase curves to constrain the radius, the albedo, and the orbital inclination of a nontransiting planet with no atmosphere and on a 1:1 spin orbit resonance.}
  {We modeled the thermal emission of a synchronous rocky planet with no atmosphere and its apparent variation with the orbital phase for a given orbital inclination. We assume that the planet is detected by radial velocity so its orbital period and minimum mass are known. We simulated observed noisy phase curves and then applied an optimization procedure to retrieve the radius and albedo of the planet and the inclination of the orbit.  }
  {Airless planets can be distinguished from planets having a dense atmosphere and their radius, albedo, and inclination (and therefore true mass) can be retrieved from multiband observations with MIRI-JWST and EChO in the 5-15~$\mu$m range. The accuracy depends on stellar type, orbital distance, radius of the planet and inclination: hot and large planets on highly inclined orbit are favored. As inclination above 60$^{\circ}$ represents half of the randomly oriented orbits, the growing population of  short-period, terrestrial-sized planets detected by radial velocity surveys and transits should offer several nearby promising targets for this method, including planets GJ581 \,e, b, and HD40307\,b.}
  {Stellar activity is likely to limit the accuracy of this method, at least for some stars. It has not been taken into account in this study, and its effects will have to be addressed in future works.}

  \keywords{Methods: numerical, statistical -- Infrared: planetary systems -- techniques: spectroscopic}
	
  \maketitle
%

\section{Introduction}
Results from Kepler \citep{Howard2011} reveal that 20-25~\% of M0 through K dwarfs host a short-period planet ($P<50$~days) with a radius between 2 and 4~R$_{\oplus}$. For the same range of orbital periods, the same study shows that more than $18\%$ of G-K dwarfs host a planet with a radius between 1 and 2~R$_{\oplus}$, which represents a lower limit, knowing that some transits are not detected for such small planets. Radial velocity surveys of F to M0 stars with HARPS have also unveiled a similar population: 
the debiased occurrence of exoplanets with M$\sin i < 30$~M$_{\oplus}$ decreases from $\sim 45\%$ at a period of ten days to $\sim 10\%$ at two days \citep{Mayor2011}. At least two thirds of this population consist of planets with M$\sin i < 10$~M$_{\oplus}$.

For the hottest of these exoplanets (e.g. Kepler 10b and Corot 7b), the geometric transit probability can reach 30\,\%, but it remains below 10~\% for most of them. As a consequence, the characterization techniques requiring transit configuration miss more than 90\,\% of these planets. This becomes a critical point when exploring the vicinity of the Sun.  Statistically, transiting planets from this population are unlikely to be found within 10~pc for K-G stars and 6~pc for M dwarfs \citep{Belu2011}. In the long run, nearby nontransiting planets and, in particular, the potentially habitable ones will be studied by direct detection with instrument derived from pioneer projects such as Darwin \citep{Cockell2009}, TPF-I \citep{Lawson2007}, TPF-C \citep{Levine2009}, or New Worlds \citep{Cash2009}, but hot planets are too close to their star for coronographs, nullers and occulters. However, we can study them by observing in multiple spectral bands the variations in the unresolved star+planet flux due to the changing phase of the planet, assuming that intrinsic stellar variations can be distinguished from the planetary modulation. Phase curves of gas giants have been observed for transiting configuration \citep[see for instance][]{Knutson2009}, but can also be measured in a nontransiting configuration \citep{Crossfield2010}. \citet{Cowan2007} measured phase variations for both transiting and nontransiting systems, and used these to constrain the inclination, albedo and recirculation efficiency. Bathala et al. \citeyearpar{Batalha2011} have measured both the phase curve and the secondary eclipse of the terrestrial planet Kepler 10b. In this case, the Kepler's spectral band (0.4-0.9~$\mu$m) means that the planetary flux measured is dominated by reflected light while the present paper addresses the thermal phase curve. Nevertheless, it is interesting to note that it is already possible to measure a phase curve with an amplitude of about 5~ppm and also that the amplitude of the phase curve is clearly better characterized than the depth of the eclipse (see Fig. 13 of Bathala et al.), even though these two quantities are equal for a synchronized transiting planet with no atmosphere  (for a nightside temperature of 0K). The precision on the eclipse depth is indeed limited by the eclipse duration, which is about 9\% of the orbital period.

The phase- and wavelength- dependent emission of hot Jupiters has been simulated with various types of atmospheric models \citep{Barman2005, Burrows2010, Fortney2006, Showman2009}. The phase-dependent visible flux and its dependence on inclination has been studied for nontransiting hot Jupiters \citep{Kane2011}. \citet{Cowan2011} studied the thermal phase curve of hot eccentric Jupiters with a semi-analytic model. The possibility of characterizing the atmosphere of a nontransiting terrestrial exoplanet by its thermal phase curves has been studied by Selsis et al. \citeyearpar{Selsis2011}.

In the present article we focus on the specific case of a rocky planet that has no atmosphere and is tidally locked in a 1:1 spin-orbit resonance on a circular orbit. Both radius and mass have been measured for some of the smallest known transiting planets (Corot~7b, GJ1214~b, Kepler~10b, planets in the Kepler~11 system). Obtained values already point to a broad diversity of objects in terms of mass-radius relationship, some probably being volatile-rich (like GJ1214~b), while some (like Kepler~10b) seem to be dominated by denser material. It is reasonable to assume that large analogs of Mercury with no atmosphere exist within this population, either because they formed dry in the hot inner part of the protoplanetary disk or because they are too hot and receive too much X/EUV irradiation to keep an atmosphere. In the case of a synchronous planet on a circular orbit, the temperature map of the planet is constant and only depends on the surface bolometric albedo. The thermal emission measured by a distant observer can thus be robustly calculated and depends on the planet radius and albedo, the orbital inclination, as well as the type of the host star and the distance the planet is from the star. For orbital inclinations over 60$^{\circ}$ (half of randomly oriented orbits), the amplitude of the phase curve is 87~\% of the secondary eclipse depth. For orbital inclinations greater than 30$^{\circ}$ (87~\% of randomly oriented orbits), the amplitude of the phase curve is 50~\% of the secondary eclipse depth. Therefore, the magnitude of the phase modulation is in most cases comparable to the occultation depth but without the restriction of the occultation duration that limits the signal-to-noise ratio. 

The present work addresses the possibility of constraining the radius, the Bond albedo, and the orbit inclination using the thermal phase curve observed in multiple spectral bands by the James Webb Space Telescope - JWST \citep{Gardner2006} or the Exoplanet Characterization Observatory - EChO \citep{Tinetti2011}, recently selected for the Assessment Phase for the Cosmic Vision program of ESA.


\section{Model}
We model the thermal emission of a planet with no atmosphere, on a circular orbit and tidally-locked in a 1:1 spin-orbit resonance. We assume a null obliquity, which is consistent with the tidal evolution of short-period planets \citep{Leconte2010}. We also assume a stable host star (see Sect. \ref{variability} for further discussions). Because insolation is constant at a given location on the planet, the surface temperature distribution only derives from the local radiative equilibrium and does not depend on the thermal inertia of the surface. Local temperatures are simply calculated by 
 \begin{eqnarray}
T(\theta)=\left(\frac{\cos(\theta)F_*(1-A)}{\sigma}\right)^{\frac{1}{4}},
 \end{eqnarray}
 where $F_*$ is the stellar flux at the orbital distance of the planet (constant for a circular orbit), and $\theta$ the zenith angle of the point source star. Considering an extended star changes the temperature near the planet terminator \citep[see Fig. 4 of] []{Leger2011} by producing a penumbra. The impact on the global emission of the planet and its apparent variations is, however, negligible even in the most extreme cases like Corot~7b, where the stellar angular diameter is  $28^{\circ}$. This is because the global emission is dominated by the hottest substellar regions, which are not significantly affected. $A$ is the bolometric surface albedo (assumed to be uniform and independent of the incidence angle, in which case it is equal to the Bond albedo of the planet), and $\sigma$ is the Stefan-Boltzmann constant.
We neglect geothermal flux, which implies that the night-side equilibrium temperature of the planet is zero. We can also set a geothermal flux $\Phi$ that adds up to the insolation and insures a minimum temperature $T_{min}=(\Phi/\sigma)^{1/4}$. This, however, affects the global thermal emission in an observable way only for unrealistically high values of $\Phi$. 

The model calculates the infrared flux emitted by the planet in any direction, assuming a blackbody emission (the emissivity of the surface is set to 1) and an isotropic distribution of the specific intensities. We use a 36 $\times$ 18 (typically) longitude-latitude grid, a resolution that is sufficient to produce smooth phase curves. The disk-integrated emission at a distance $d$ is given by 
 \begin{eqnarray}
F_\lambda=\left(\frac{R}{d}\right)^2 \sum_j I_\lambda(T_j)S_j \cos(\alpha_j)
 \end{eqnarray}
 where 
$R$ is the radius of the planet, $I_\lambda(T_j)$ is the blackbody specific intensity (W m$^{-2}$ $\mu$m$^{-1}$ ster$^{-1}$) at the temperature $T_j(\theta)$ and at the wavelength $\lambda$, $S_j$ is the surface of the cell $j$, and $\alpha_j$ is the angle between the normal to the cell and the direction toward the observer. Only locations visible to the observer ($\cos\alpha_j> 0$) contribute to the sum. Because the orbit is circular, the observing geometry is only defined by the orbit inclination $i$.

\begin{figure*}
\begin{center}
\includegraphics[width=\linewidth]{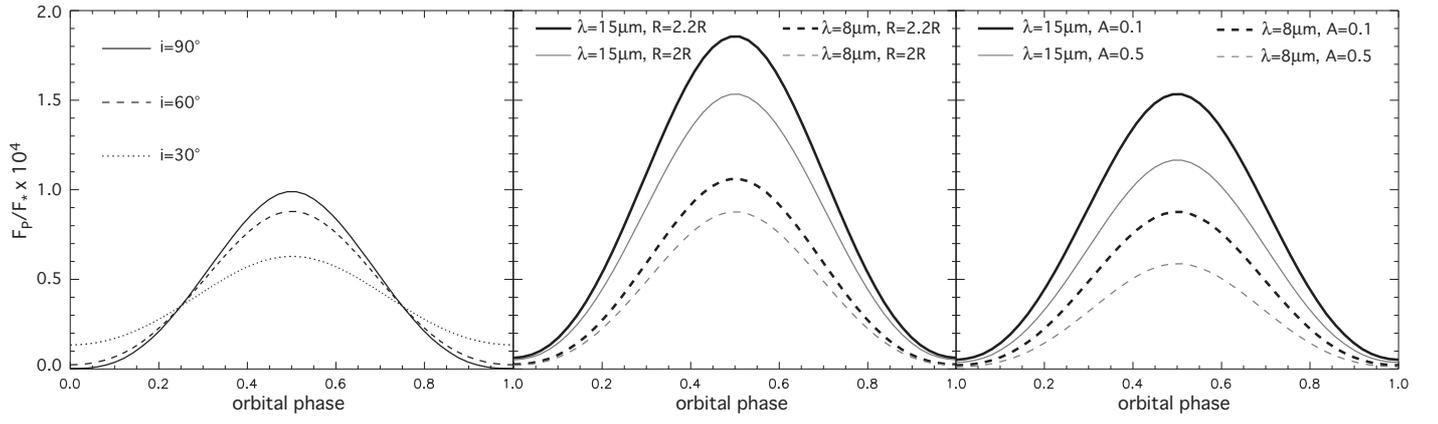}
\end{center}
\caption{Effect of inclination (left, $R=2\,R_{\oplus}$ and $A=0.1$), radius (middle, $i=60^{\circ}$ and $A=0.1$) and albedo (right, $i=60^{\circ}$ and $R=2\,R_{\oplus}$) on the flux ratio, at 8 (default) and 15~$\mu$m. Phase 0 corresponds to the minimum fraction of the illuminated planet received by the observer (0 if $i=90^{\circ}$). The eclipse normally occurring for $i=90^{\circ}$ is not shown here.}
\label{phasecurves3}
\end{figure*}

\begin{figure}[ht]
\begin{center}
\includegraphics[width=\figwidth]{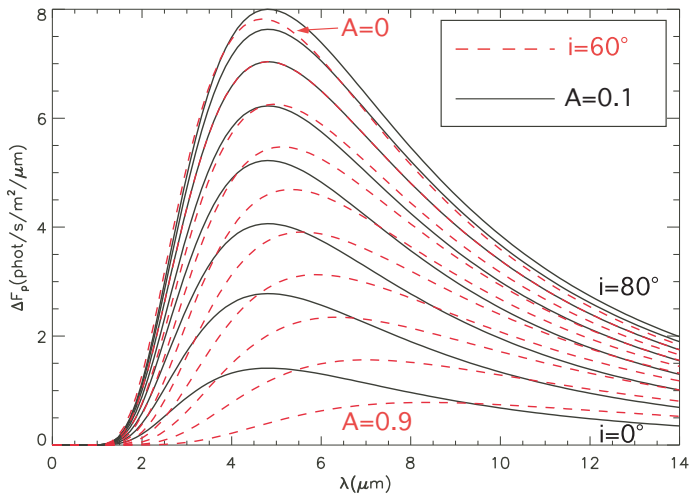}
\end{center}
\caption{Effect of inclination and albedo on the variation spectrum. The peak amplitude of the phase curves is given as a function of wavelength for a planet at 0.04~AU from a 0.5~M$_\odot$ star. Solid curves are obtained with an albedo of 0.1 and inclinations ranging from 0 to 80$^{\circ}$. Dotted curves are calculated for an inclination of 60$^{\circ}$ and albedo ranging from 0.1 to 0.9.
}
\label{spec}
\end{figure}

\begin{figure}[p]
\begin{center}
\includegraphics[width=\figwidth]{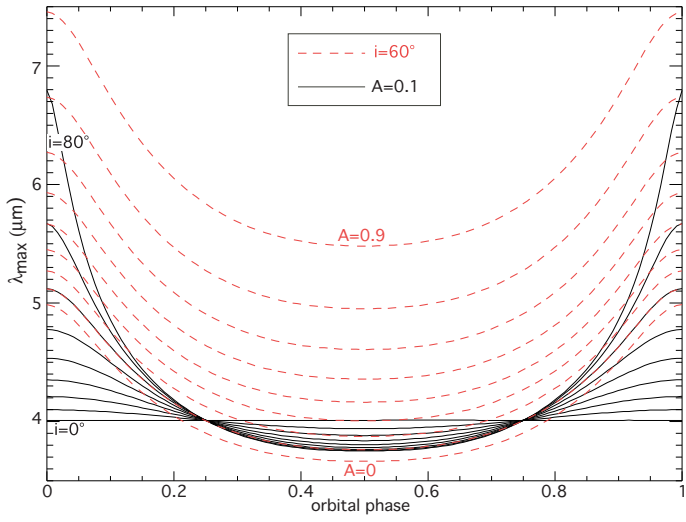}
\end{center}
\caption{Influence of inclination and albedo on the wavelength $\lambda_{max}$ of the spectrum peak.  Calculations are done for a planet at 0.04 AU from a 0.5~M$_\odot$ star, for inclination between 0 and 80$^{\circ}$ and an albedo of 0.1 (solid lines) and for an albedo between 0 and 0.9 and an inclination of 60$^{\circ}$ (dashed). }
\label{A_i}
\end{figure}

Figure~\ref{phasecurves3} presents thermal phase curves obtained with the model. We can see the influence of inclination, radius, and albedo for two wavelengths. In a nontransiting configuration, photometry will not yield the absolute value of the planetary flux (which can be inferred from secondary eclipses), but only its variation with the orbital phase. Figure~\ref{spec} shows the peak amplitude of the modulation as a function of wavelength, called the \textit{variation spectrum} by Selsis et al. \citeyearpar{Selsis2011}. Phase curves obtained in one single spectral band can be reproduced by various combinations of $R$, $A$, and $i$, but multiwavelengths measurements break this degeneracy. The disk-integrated thermal emission simply scales as $R^2$ at all wavelengths, while $A$ and $i$ both have an effect on the spectral distribution. The albedo directly controls the intrinsic spectral properties of the disk-integrated planetary flux through the distribution of surface temperatures, and thus determines the wavelength of variation-spectrum maximum. The inclination, on the other hand, controls the amplitude of the phase variations seen by the observer. This effect of the inclination is also wavelength-dependent, but it has a negligible effect on the position of the variation-spectrum maximum as seen in Fig.~\ref{spec}. Figure~\ref{A_i} presents this effect in a different way by showing how changing the albedo or the inclination shifts wavelength of the emission maximum ($\lambda_{max}$) as a function of the orbital phase.

It is therefore possible, in theory, to infer the radius, the albedo, and the inclination of a planet from the phase variations of its thermal emission, at least with noise-free observations.
\section{Methodology}
\subsection{Producing a phase curve}
We consider a nontransiting exoplanet on a circular orbit (found for instance by radial velocity surveys). Some characteristics of the system are known to the observer: its distance, the stellar properties, the orbital period of the planet, and the ephemeris. The planetary radius, albedo, and the orbital inclination (as well as the true mass), on the other hand, are unconstrained. The question we address is the ability to retrieve these three characteristics using spatially-unresolved spectro-photometric observations covering one (or more) orbital period. With our model, we can produce these thermal phase curve variations and add unavoidable noise to them. To calculate the stellar photon noise, we need to assume a collecting area, a data point binning time, a spectral resolution and a target distance (here set to 10~pc). We neglect here other stellar phenomena such as stellar spots, flares or pulsations (see Sect. \ref{variability}). We assume that one complete orbit of the planet is sampled in 20 data bins of equal duration. We considered the collecting areas of the James Webb Space Telescope (25~m$^2$) and of EChO (1.1~m$^2$). We consider ten 1 micron-wide spectral bins from 5 to 15~$\mu$m. We use this for both the Mid InfraRed Instrument (MIRI) on the JWST \citep{Wright2004}, and EChO spectrometers. We have similar results for the MIRI filters with three broad bands centered on 7.7, 10.0, and 12.8 $\mu$m, with widths of 2.2, 2.0 and 2.4 $\mu$m, respectively, corresponding to the filters of MIRI. This tells us that little spectral resolution is enough to break degeneracies.

The considered MIRI spectrometer provides spectroscopic measurements over the wavelength range 5 - 28.3~$\mu$m, but the thermal noise of the telescope dominates above 15 $\mu$m. For the EChO spectrometers, the measurements could actually cover the 0.5-16 $\mu$m range, and integrating to wavelengths shorter than 5~$\mu$m could improve the results, especially for hot planets (see Sect. \ref{Discussion}), but we kept this $5-15$~$\mu$m range for comparison purposes. In addition, the planet/star contrast ratio drops below 10~ppm at wavelengths shorter than 5~$\mu$m for the coldest planet we consider (see Sect. \ref{Results}). A photometric precision better than 10~ppm, even for several days of integration may not be achievable for the forthcoming generation of infrared telescopes. Furthermore, the reflected light has to be taken into account at short wavelengths (below 3~$\mu$m) for the planets we consider. Our model uses the isotropic assumption for the planetary flux, which is a fair approximation for thermal emission but less valid for the scattered light (see Sect.~\ref{lambert}). Several additional parameters would need to be added to model the reflected signal correctly. The associated degeneracies and the wavelengthdependence of the albedo are such that extending the wavelengths range to reflected light is not likely to improve the retrieval.
For given albedo, inclination, and radius, we produce $n$ noisy phase curves for each spectral band. The only difference between these $n$ realizations is the random stellar photon noise, and instrumental noise when included. For each noisy phase curve, we determine the values of $R$, $A$, and $i$ that minimize $\chi^2$ using the downhill simplex method. From these $n$ different estimates of $R$, $A$, and $i$, we determine their median value and the associated error defined as the smallest interval containing 95~\% of the retrieved values (the ``2-$\sigma$'' confidence level if the values have a normal distribution). With zero eccentricity and obliquity, we only have these three parameters to constrain, but our method is applicable to more general configurations (not addressed in the present study), requiring the retrieval of more parameters. This is, for instance, the case with nonsynchronized oblique planets for which four additional unknown parameters control the thermal emission: the rotation period, two angles for the rotation vector, and the surface thermal inertia. 

 \subsection{The instrumental noise} 
 \label{noise}
 In Belu et al. (2011) we showed that, on average, taking instrumental noises into account divides the signal-to-noise ratio (S/N) by two compared with shot noise, for terrestrial-exoplanet transit spectroscopy and photometry with JWST. We therefore replicate here the modeling in this reference for both EChO and JWST/MIRI: read-out noise, dark current, optics thermal emission, and reduction of exposure time due to readout rate. We do not take the zodiacal contribution into account, both domestic and at the target system, because its variations are not likely to be in the same regime of frequencies as the signals considered here. Modeling jitter noise \citep{Deming2009} is beyond the scope of the present work, but in-flight calibrations may very well be able to mitigate this effect \citep[see][for latest methods and performances]{Ballard2010}. The exact calculations are detailed very well in the program used to compute them, which is available on demand.
 
 Given the broad spectral range aimed at by EChO, it will require being divided into several sub-bands. Our $5-15\mu$m band will be divided into two sub-bands (Table~\ref{detectors-EChO}, \citealp{Tinetti2011}; Marc Ollivier, priv. comm.). It has not yet been decided which detectors set (all-Si:As or MCT+Si:As) will be used for EChO. However, in our simulations, we consider only the option of all-Si:As detectors, since the other one gives comparable results. The characteristics of JWST/MIRI are readout noise RMS 19~e$^-$/pixel, dark current 0.03~e$^-$/s/pixel, and a full-well capacity $10^5$~e$^-$.
    
 \section{Results}
 \label{Results}
\begin{figure*}[p]
\begin{center}
\includegraphics[width=\linewidth]{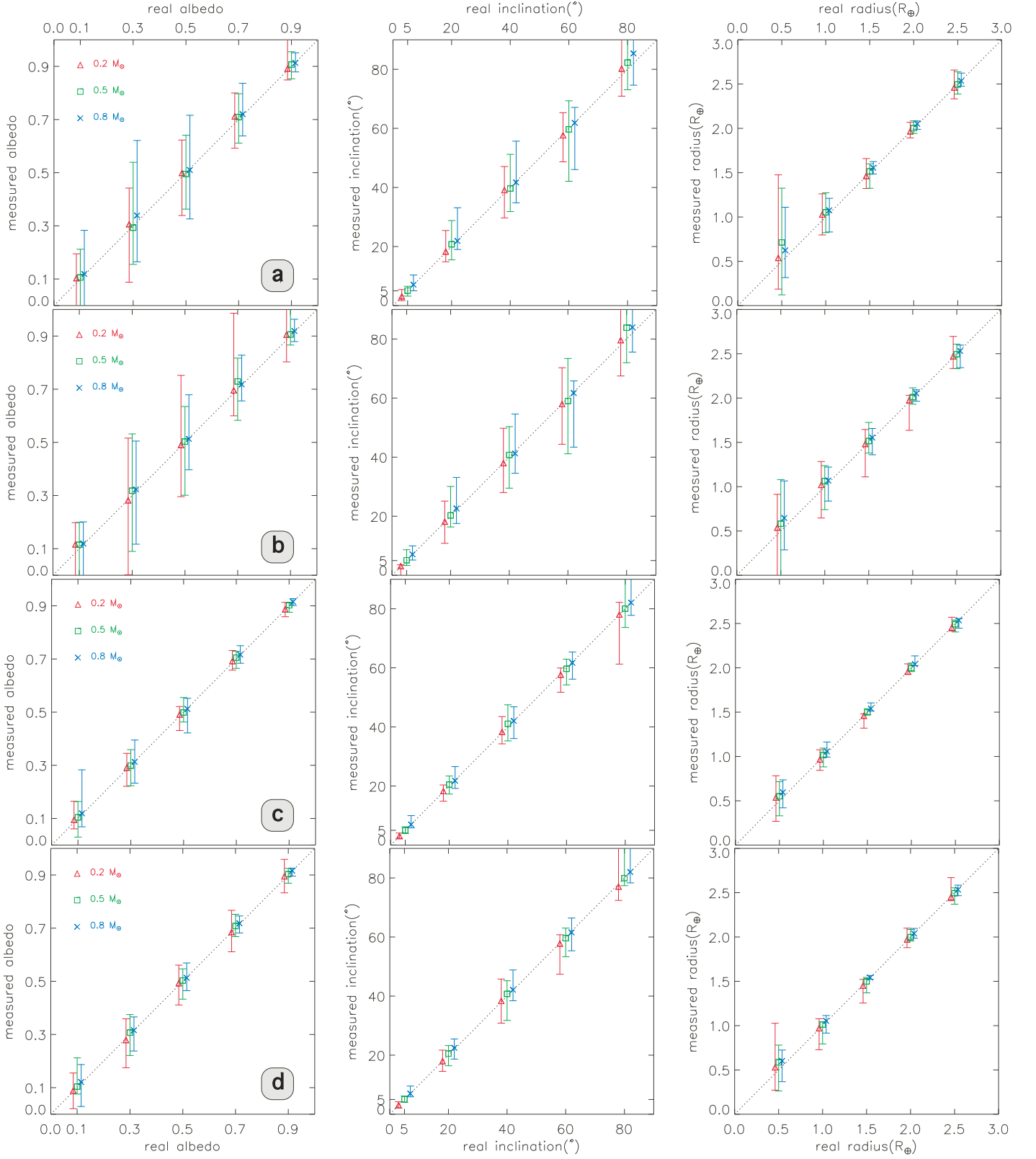}
\end{center}
\caption{Ability to retrieve the albedo, inclination and radius of a synchronous rocky exoplanet with a 1.2~m telescope like EChO, whose instrumental noise is simulated with Si:As detectors (rows: a, b) and with the MIRI/JWST (c, d) for different stellar masses. The orbital distance is $a$=0.02\,AU (a, c) and $a$=0.04\,AU (b, d). The triangles, squares, and crosses correspond to 3 different stars ($M$=0.2, 0.5, and 0.8\,$M_\odot$ respectively). Error bar include 95~\% of retrieved values. The median of the retrieved values is indicated inside the error bar. }
\label{all_results}
\end{figure*}

\begin{figure}[p]
\begin{center}
\includegraphics[width=\figwidth]{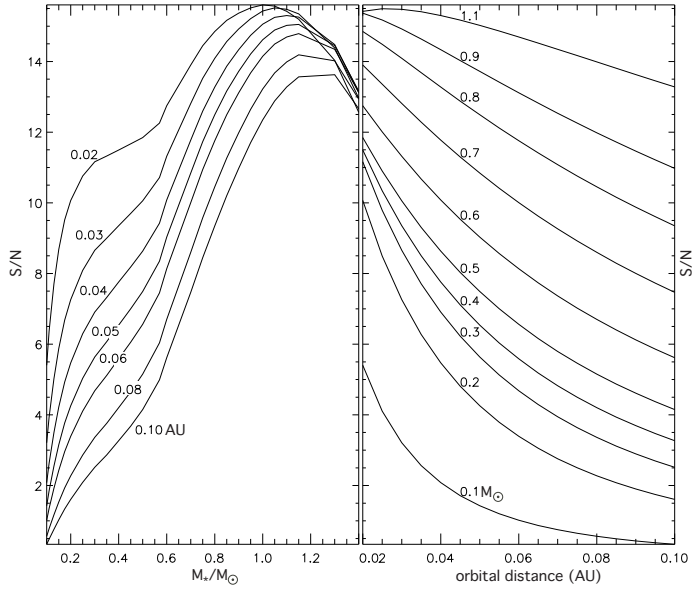}
\end{center}
\caption{Estimated signal-to-photon-noise ratio (S/N) as a function of stellar mass and orbital period. Each curve in the left plot is for one orbital distance, and each curve on the right plot is for one stellar mass. Here, we consider the planetary signal to be the peak amplitude of the phase curve, integrated from 5 to 15~$\mu$m intervals. The duration of the two data bins used to compute the peak amplitude is set to 1/20$^{th}$ of the orbital period, the inclination to 60$^{\circ}$, the distance to 10~pc, and the telescope diameter to 1.5~m telescope.}
\label{snr_p_mstar}
\end{figure}

\begin{figure*}[p]
\begin{center}
\includegraphics[width=\linewidth]{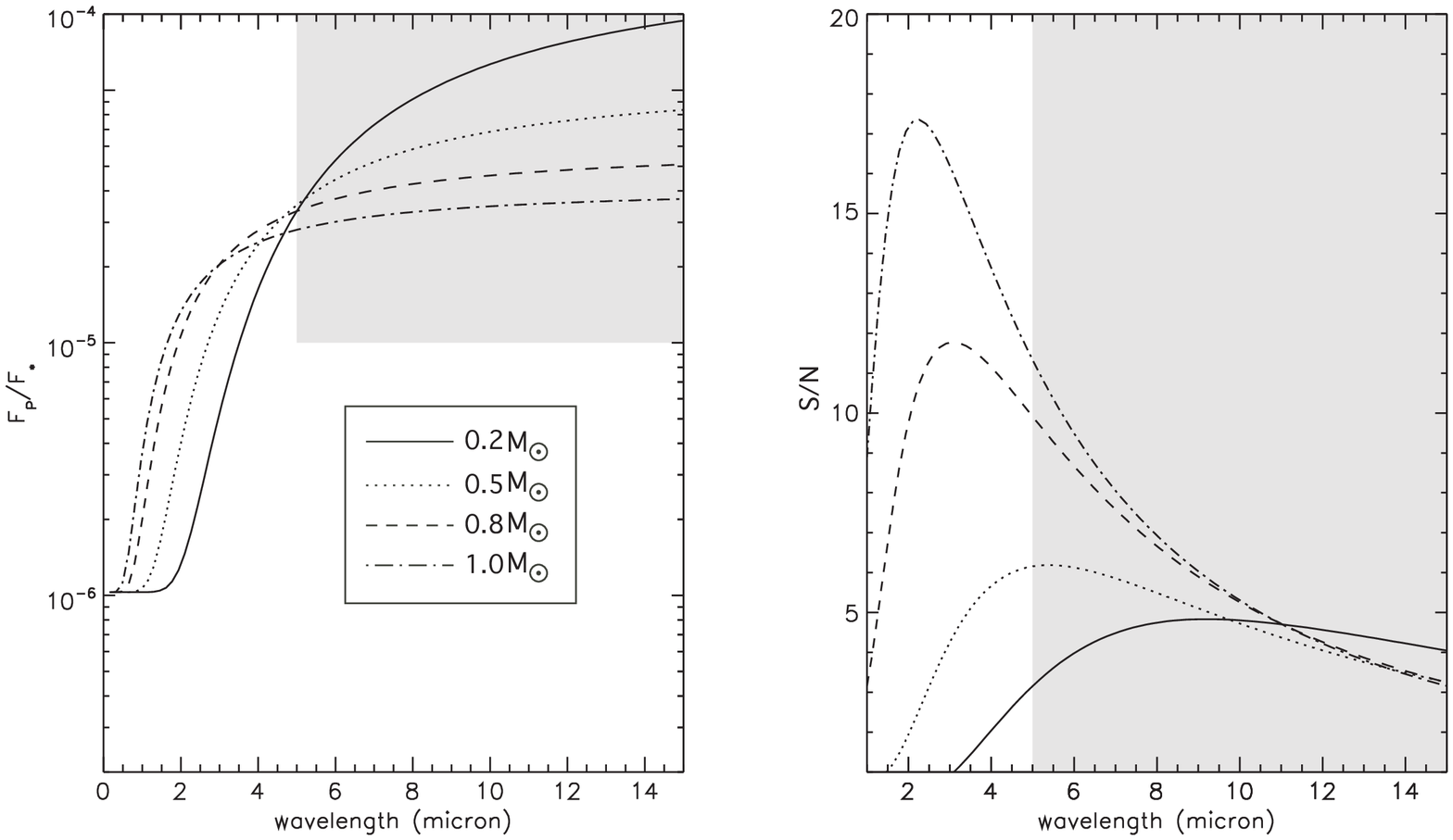}
\end{center}
\caption{Variation spectrum as planet/star contrast (left) and S/N (right). These plots give the peak amplitude of the phase variation as a function of wavelength. The orbital distance of the planet is 0.02~AU, its radius 2~R$_{\oplus}$, and its inclination $60^{\circ}$. Each curve is for a given stellar mass. To calculate the S/N, we set $\Delta \lambda=1~\mu$m, distance=10~pc, telescope diameter=$1.5$~m. The gray area corresponds to the $5-15\mu$m range (left \& right) and to contrasts higher than $10^{-5}$ (left).}
\label{contrast_vs_snr002}

\begin{center}
\includegraphics[width=\linewidth]{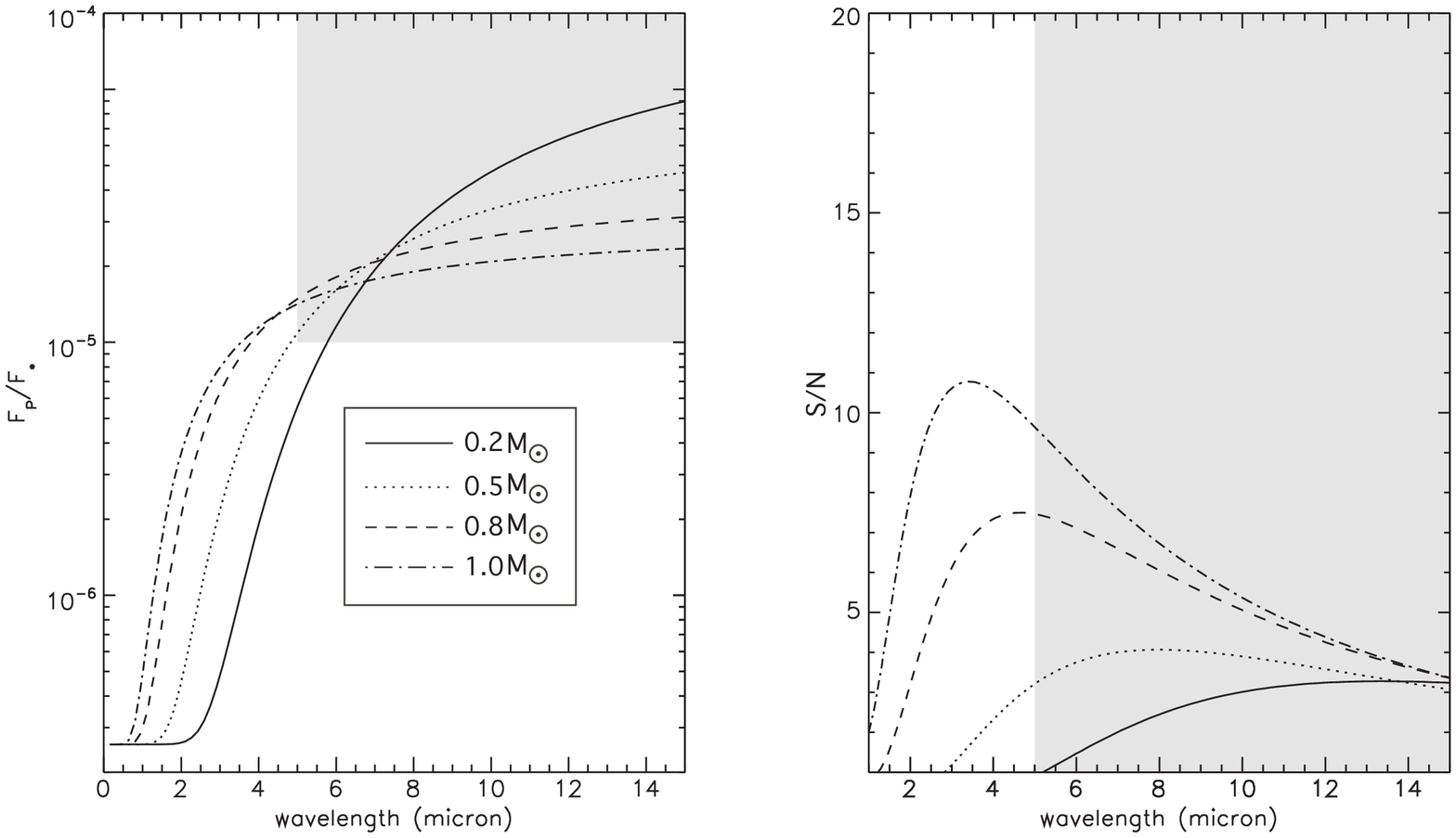}
\end{center}
\caption{Same as Fig.~\ref{contrast_vs_snr002} for an orbital distance of 0.04~AU.}
\label{contrast_vs_snr004}
\end{figure*}

The properties of the studied star+planet systems are summarized in Table~\ref{cases}, which gives the stellar masses, orbital distances (and periods) and corresponding equilibrium temperatures of the planets (ranging from 420 to 1640~K). The stellar masses represent the stars of the solar neighborhood (dominated by M and K stars. The range of orbital distances is chosen to offer a high S/N. We test our procedure to retrieve the radius, albedo, and inclination in different cases and with different instruments. 
Figure~\ref{all_results} shows the accuracy in retrieving the radius, the albedo, and the inclination of a planet for two instruments (EChO and JWST/MIRI), for different stellar types and for orbital distances. In all these plots, the X-axis gives the \textit{real} values of $R$, $A$, and $i$, which are used to produce the noisy phase curves, and the Y-axis gives the median of the best-fit values. The error bar gives the 95~\% confidence level ($2~\sigma$). A good retrieval should fall on the dotted line. For the albedo retrieval (left), the radius is fixed to 2\,$R_{\oplus}$ and the inclination to $60^{\circ}$, which is the median value for randomly oriented systems. For the inclination retrieval (middle), the radius is fixed to 2\,$R_{\oplus}$ and the albedo to 0.1. For the radius retrieval (right), the albedo is fixed to 0.1 and the inclination to $60^{\circ}$. All plots include stellar photon noise and instrumental noise.
The accuracy on the retrieved parameters depends both on the S/N that can be achieved in measuring the variations in the planetary signal, but also in the shape of the spectral distribution of the planet emission in the observation bands, which allows us to break the degeneracy between radius and temperature. Figure~\ref{snr_p_mstar} shows how the S/N depends on the stellar mass and the orbital period. For this plot, the S/N is calculated as follows. We integrate this variation spectrum over the $5-15$~$\mu$m range and compare this signal with the photon noise of the star integrated over the same wavelength range. One can see that the S/N monotonically decreases as the orbital distance increases and monotonically increases with the stellar mass between 0.1 and 1~M$_{\odot}$. This trend generally explains the increase or decrease in the error bars when changing the orbital period or the stellar mass. This is, however, not always true because the ability to retrieve $A$, $R$, and $i$ also requires measuring the wavelength dependency of the spectra. An important factor is thus the position of the emission peak compared with the observation spectral window. This explains the turnover around 1$M_\odot$. An appropriate window can compensate for lower S/N and vice versa. As an extreme example, we consider a very hot planet observed at long wavelengths where the emission of the substellar area (dominating the phase curve) is in the Rayleigh-Jeans regime. In this case, despite a possibly very high S/N, the weak wavelength-dependency would result in an enhanced degeneracy between radius and temperature. This effect can be seen, for instance, on the albedo and inclination retrieval in Fig.~\ref{all_results}a: error bars can be similar to or smaller for the less massive star or at larger orbital distance, contrary to what S/N suggests. This is because cooler planets have an emission peak closer to or inside the $5-15~\mu$m domain and exhibit a better temperature signature. The position of the variation-spectrum peak can be seen in Figs.~\ref{contrast_vs_snr002} and \ref{contrast_vs_snr004}. This tells us that optimal retrieval will be achieved when the observation window includes the shortest wavelength not affected by reflection. In this article we did not adapt the window to the planet properties, so our retrieval is thus not optimal.

For inclinations close to 90$^{\circ}$, the flux becomes roughly proportional to $\sin$~$i$, hence weakly sensitive to $i$. For this reason, inclinations in the $80-90^{\circ}$ range cannot always be distinguished, and the uncertainty on the inclination increases with the inclination despite larger modulations.

Although the accuracy obtained with EChO is lower than JWST for a given number of orbits, EChO will have the possibility of dedicating more time to a given target, reducing the S/N to levels comparable to or better than those achievable with JWST, hopefully with a significantly improved stability and the capability of observing spectral phase curves over a wider spectral domain simultaneously. Using the $2.5-5~\mu$m domain would better characterize the hottest planets. On JWST, measurements in this window imply the use of NIRSpec, an instrument that cannot be used simultaneously with MIRI, while EChO spectrometer should cover the whole $0.5-16~\mu$m window.

The results presented in this work assume that the observation covers two orbital periods, which will certainly be required in practice to extract the periodic planetary signal from the stellar variability. We tested on several cases that the error on the albedo, the inclination and the radius decreases as $1/\sqrt{N}$, where $N$ is the number of orbits observed.  Also for observations covering $N$ orbits, the values of $R$, $A$, and $i$ obtained for each individual orbit (or for a number of orbits smaller than $N$) provide an information on their dispersion.

\subsection{Constraining both mass and radius}
\begin{figure*}[p]
\begin{center}
\includegraphics[width=\linewidth]{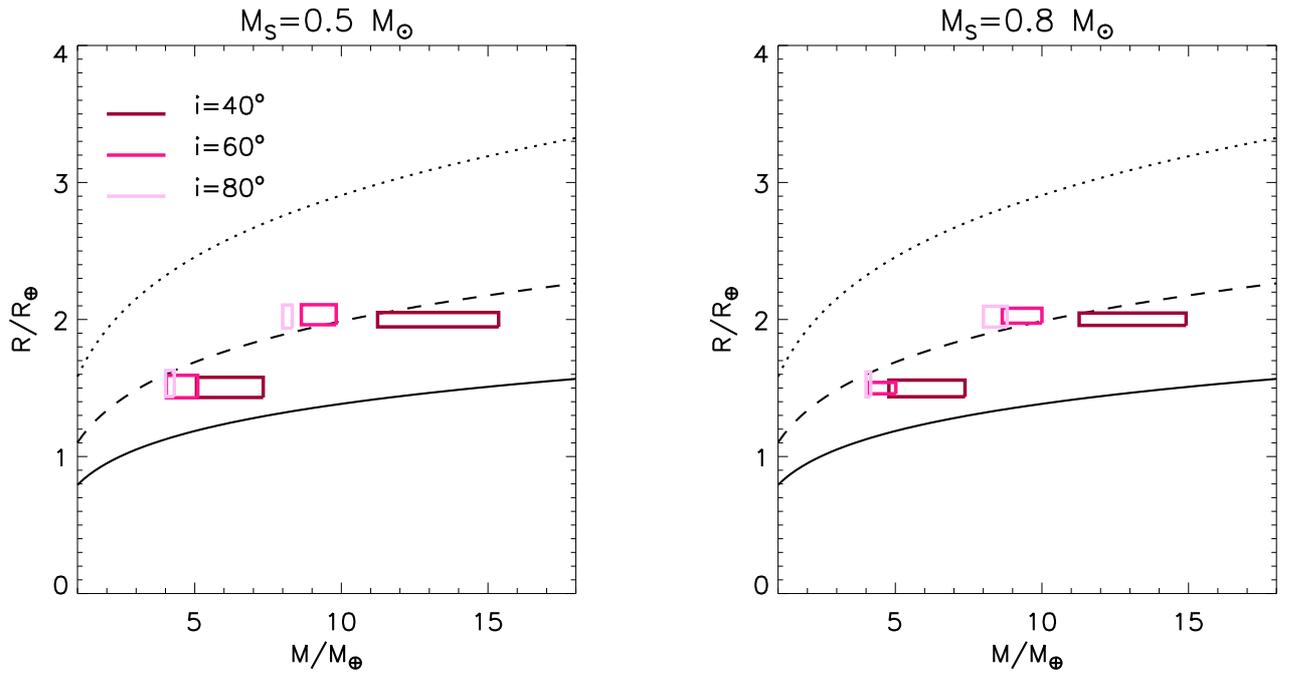}
\end{center}
\caption{Constraining the bulk composition.  Boxes indicate the 95~\% confidence level on the retrieval of both the radius and the mass for two synchronous planets ($R=1.5~R_{\oplus}$, $M\sin{i}=4~M_{\oplus}$, and $R=2~R_{\oplus}$, $M\sin{i}=8~M_{\oplus}$), at 0.02\,AU of a 0.5 and a 0.8~$M_{\odot}$ star, observed with the JWST for 3 different inclinations (2 orbits for computation). The solid, dashed and dotted lines correspond respectively to a pure iron, pure rock, and pure-ice planet according to Fortney et al. \citeyearpar{Fortney2007err}. }
\label{masserad}
\end{figure*}

If we combine the projected mass $M \sin i$ measured from radial velocity observations with the constraint on the inclination from the phase curve, we obtain an estimate of the true mass. Retrieved mass and radius and their associated uncertainty can be compared with theoretical models to assess the composition of the planet. Figure~\ref{masserad} shows mass-radius relations of ice/rock and rock/iron planets from \citet{Fortney2007err} and the range of mass and radii obtained from phase curves for two planets ($R=1.5~R_{\oplus}$, $M\sin{i}=4~M_{\oplus}$ and $R=2~R_{\oplus}$, $M\sin{i}=8~M_{\oplus}$) around a 0.5 and a 0.8~$M_{\odot}$ star with an orbital period of three days. The planets around the 0.8~$M_{\odot}$ star logically give the best estimate of the composition. The uncertainty on the composition is dominated by the uncertainty on the mass, and one should also include the error on the measurement of $M \sin{i}$ itself. Interestingly, the uncertainty on the planet radius does not come directly from the uncertainty on the stellar radius as it does when the planetary radius is inferred from the primary transit depth, in the case of transiting planets. Indeed, the depth of the transit provides the ratio $R_{p}/R_{\star}$ and an estimate of $R_p$ therefore requires a value for $R_{\star}$, which comes with its uncertainties. When deduced from the thermal phase curve, the radius estimate is also affected by uncertainties (for instance on the luminosity of the star or on the orbital distance), but not coming directly from the radius. Errors on the stellar luminosity and the stellar mass (used to convert an orbital period into an orbital distance) do, however, result in retrieval errors.

\begin{figure}[p]
\begin{center}
\includegraphics[width=\figwidth]{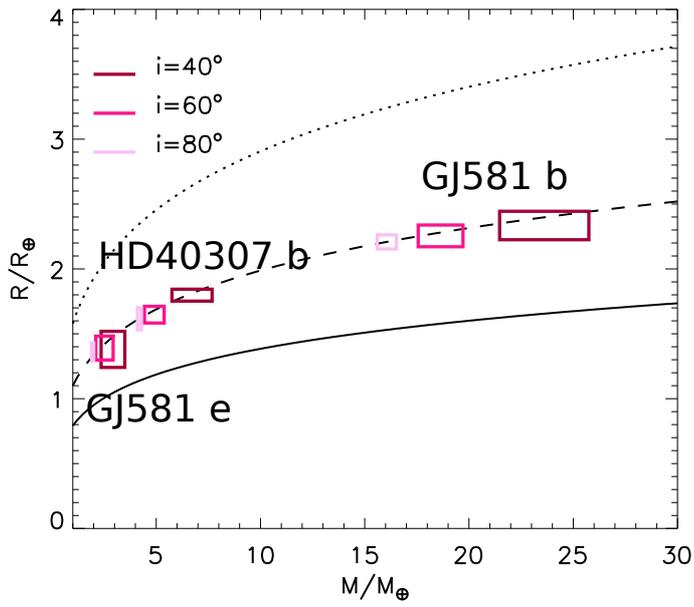}
\end{center}
\caption{Precision of the mass and radius determination for three known exoplanets (GJ581~b, GJ581~e, and HD40307~b). Similar to Fig.~\ref{masserad}, but we no longer assume a fixed radius for the planets, but a rocky composition (100~\% silicates). The radius changes with the mass and thus with the inclination.}
\label{MR-candidates}
\end{figure}

As another illustration, we show in Fig.~\ref{MR-candidates} the result for three known exoplanets (HD40307\,b, GJ581 b, and GJ581 e, see Sect.~\ref{candidates}), assuming that they are made of silicates (an assumption that is probably not realistic for GJ581 b, whose volatile content must be high considering its large mass).

\subsection{Effect of an atmosphere}
We do not know a priori whether the observed planet has an atmosphere of not. As shown by Selsis et al. (2011), the presence of an atmosphere can be inferred from the variation spectrum. We did, however, test our $R-A-i$ retrieval procedure (which assumes no atmosphere) on phase curves computed for a planet with a dense atmosphere. The planet is a 1.8~R$_{\oplus}$ rocky planet with a 1~bar CO$_2$ atmosphere, in a 1:1 spin-orbit resonance on an 8-day orbit around an M3 dwarf, with $60^{\circ}$ inclination. The structure of the atmosphere and the associated radiative transfer has been modeled with a 3D GCM (global climate model) described in \citet{Selsis2011}.

\begin{figure*}[p]
\begin{center}
\includegraphics[width=1\linewidth]{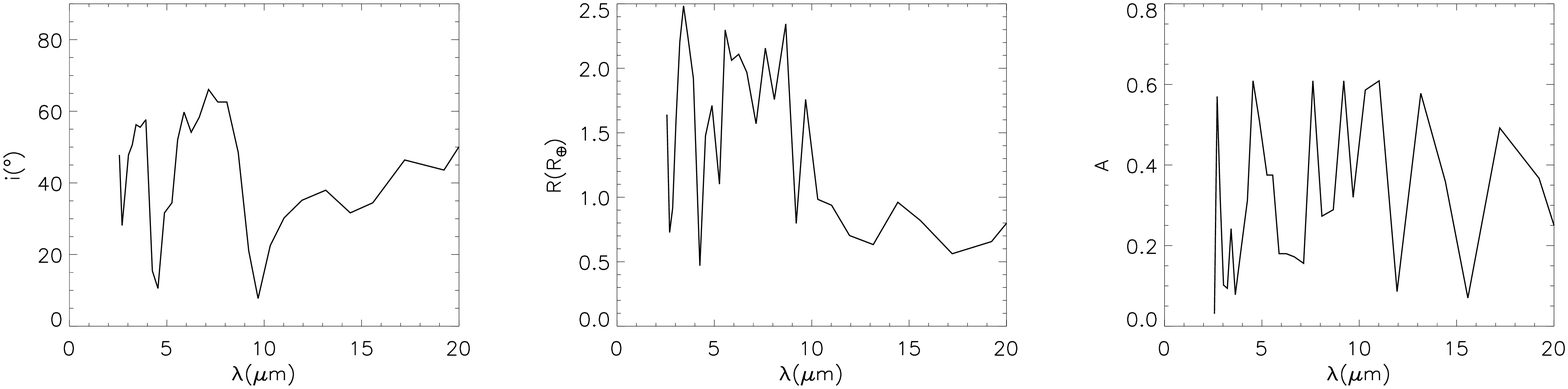}
\end{center}
\caption{Best fit parameters as a function of wavelength, for a planet with a 1 bar $\mathrm{CO_2}$ atmosphere.}
\label{specatmo1}

\begin{center}
\includegraphics[width=1\linewidth]{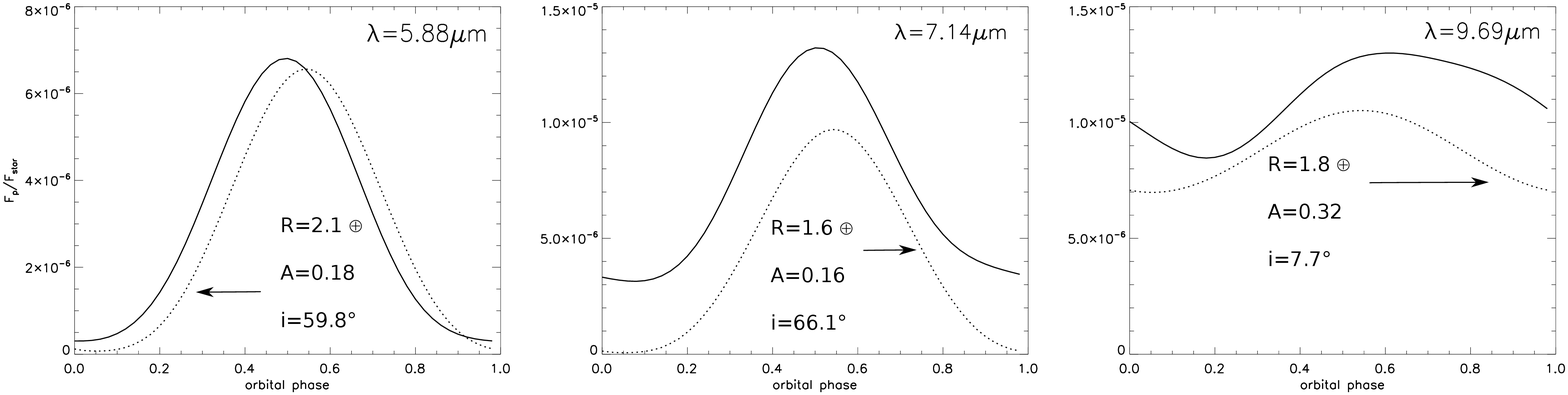}
\end{center}
\caption{Phase curves of a planet with a 1~bar $\mathrm{CO_2}$ atmosphere (solid lines) and best fit of the variation (and not of the absolute flux) obtained assuming an airless planet (dashed lines). The fit is done for each wavelength separately.}
\label{atmo}
\end{figure*}

Our model obviously fails to fit the phase curves variations at all wavelengths, which by itself shows that the ``no atmosphere'' assumption is wrong. If we treat each spectral band independently of the others, our model can obtain a set of $R$, $A$, and $i$ with reasonable $\chi^2$ at some wavelengths but the retrieved values strongly vary from one wavelength to another. Figure~\ref{specatmo1} shows the retrieved values as a function of wavelength (no noise is considered in this retrieval). As in the variation spectrum described by Selsis et al. \citeyearpar{Selsis2011}, the signatures of molecular absorption can be seen in these plots, in particular the 2.7, 4.3, and 15~$\mu$m bands of CO$_2$ in the ``inclination spectrum''. This is because different wavelengths probe different altitudes in the atmospheres, with different day-night temperature contrasts. In absorption bands, the flux comes from high altitudes with smooth day-night contrast, while in atmospheric windows, the flux comes from the surface that exhibits a temperature distribution that differs less from the airless case.

As an example, we show in Fig.~\ref{atmo} the results of the fit for three wavelengths. We only fit the variations and not the absolute flux, which is why the solution found by our procedure can produce a phase curve that is shifted vertically. Because we only consider synchronized planets, which is equivalent to a planet with no thermal inertia, our modeled phase curves cannot exhibit the phase shift that atmospheric circulation can produce. If we no longer consider synchronized planets and let the rotation rate and the thermal inertia be free parameters, our airless model may fit a monochromatic phase curve obtained with an atmosphere with better agreement, by reproducing the phase shift. In this case, the displacement of the hot spot (compared with the exact substellar location) is due not to horizontal circulation, as in an atmosphere, but to vertical heat diffusion. However, retrieved rotation rate and thermal inertia would also depend on the wavelength, as for the inclination, radius, and albedo. Also, for planets hot enough to produce an observable infrared phase curve, on circular orbits, synchronization occurs on extremely short timescales so there is no justification for considering nonsynchronized solutions. A phase shift can thus be attributed to an atmosphere alone. This point will have to be stressed more carefully for cooler planets, which are less subject to tidal forces, and which need to be observed by future direct detection techniques. 

If no photometric variations are seen despite accurate photometric measurements, we may be able to infer that the planet has a dense atmosphere, as attempted on GJ876\,b by Seager and Deming \citeyearpar{Seager2009}. The absence of modulation can also be due to an inclination close to 0$^{\circ}$, but in some cases inclinations lower than a given value can be rejected. This can be done using the measurement of the projected rotation of the star and assuming that planetary orbits remain close to the stellar equator (known to be wrong for many hot Jupiters), but also using dynamical constraints, which was done for the systems GJ876 \citep{Correia2010} and GJ581 \citep{Mayor2009a}. A lack of modulation can also be come from an extremely high albedo, which may be checked at short wavelengths.
%

\section{Discussion}
\label{Discussion}

\subsection{Validity of the radiative model and reflected light}
\label{lambert}
The ability to constrain $R$, $A$, and $i$ from spectral phase variation comes from the simplicity of the model that assumes an isotropic distribution of the thermal emission, uniform surface properties, and an emissivity independent of the wavelength and equal to 1. We now discuss the validity of these assumptions.

An emissivity value $\epsilon$ lower than 1 but independent of $\lambda$ would only mean that the ``effective albedo'' we infer is in fact equal to $(A+\epsilon -1)/\epsilon$ and does not affect the radius and inclination retrieval. Variation in the emissivity with wavelength (for instance from the 10~$\mu$m silicate band), if significant enough to affect the planetary emission in an observable way, would be seen in the variation spectrum just as an absorption feature due to an atmosphere (see Selsis et al. 2011). A correction could thus be done a posteriori (probably yielding larger uncertainties on the retrieval). The variation in the surface albedo with wavelength influences our model only in terms of emissivity as our model is only sensitive to the bolometric surface albedo (the fraction of reflected energy integrated over the whole spectrum). We tested our retrieval algorithm that assumes a uniform albedo on phase curves computed with a nonuniform albedo. To each individual cell of the surface, we gave a random albedo using a normal distribution. We did the test for a mean albedo of 0.1, 0.3,...,0.9 with a standard deviation of 0.1. Error bars on the retrieved albedo are centered on the mean value of the normal distribution, and the error bars are twice as broad as with a uniform albedo, but there is no noticeable effect on the inclination and radius retrieval. We did not test the effect of having large regions with different albedos. An interesting case to be addressed in the future is to consider a surface composition (and thus a surface albedo) changing at a given temperature (and thus incidence angle). This could be relevant for very hot planets like Corot~7b or Kepler~10b, which have a strong temperature gradient and which may have substellar lava-oceans covering a $0<\theta<45^{\circ}$ area \citep{Leger2011}.

Because of its very low thermal inertia and the length of its solar day ($\sim 27$~days), the temperature distribution on the day side of the Moon is similar to that of a synchronized planet, and brightness temperatures on the day side of the Moon follow the $\cos(\theta)^{\frac{1}{4}}$ law, where $\theta$ is the incidence angle  \citep{Lawson2000}. This means that the isotropic assumption for the emission is a good approximation (with no significant effect of roughness and craters), but also that variations on albedo with location are small enough. Light reflected by the moon does not, however, follow a Lambertian distribution (the reason the full Moon seems so ``flat'' to us). As other rough planetary surfaces with a low albedo, the diffuse reflection is described better by the Lommel-Seeliger law \citep{Fairbairn2005}. Because of the non-Lambertian behavior of the reflected light (which can also include a specular component for high albedo), we restricted our work to wavelengths ($\lambda > 5\mu$m) where the reflected component is orders of magnitude lower than the emission. However, the temperature distribution, and thus the thermal emission, can be affected by the dependence of the surface albedo on the incidence angle $\theta$ found in the Lommel-Seeliger phase function. We tested the impact of this effect on the thermal phase curve and found it to be insignificant on the phase curve. The reason is that surface temperature is affected only at high incidence angle, near the terminator where the temperature is too low to contribute significantly to the disk-integrated emission. Small departures from an isotropic behavior of the thermal emission do exist on the Moon and other planetary surfaces (due for instance to roughness and craters) but we neglect these effect in this study. Knowing that the retrieval of $R$, $A$, and $i$ is feasible, it will become necessary to address this question in further works.

\subsection{Existing candidates}
\label{candidates}
Among the published planets detected by radial velocity and at the time of writing, at least seven are potentially terrestrial objects with characteristics allowing a measurement of the phase variations with JWST or EChO\footnote{Unless their ecliptic latitude does not allow pointing.}:  a minimum mass below 20~M$_{\oplus}$, an orbital distance within 0.05~AU, and membership in a system closer than 15~pc. According to estimates \citep{Mayor2011}, more such planets remain to be found. Expected detections with 15~pc critically depend on the occurrence of planets around M stars, which is yet poorly constrained \footnote{A study by \citet{Bonfils2011} based on HARPS measurements,  actually shows that ~35~\% of M stars planets host a 1-10 M$_\oplus$ planet.}. The occurrence rate as inferred from Kepler candidates \citep{Howard2011} shows no decrease (or even a slight increase) from K to M0 dwarfs. Assuming the same frequency of exoplanets for M and K stars yields hundreds of candidates within 15~pc.  Within the current sample of seven, three have zero eccentricity (at the precision of the measurements): GJ581 \,e and b \citep{Mayor2009a}, and HD40307\,b \citep{Mayor2009b}. The other four have eccentric orbits: GJ674\,b: $e \sim 0.2$ \citep{Bonfils2007}, GJ876\,d: $e \sim 0.2$ \citep{Rivera2010}, 61Vir~b: $e \sim 0.1$ \citep{Vogt2010} and 55Cnc~e: $e \sim 0.25$ \citep{Fischer2008}. To model the phase curve of an eccentric planet requires including heat diffusion on the subsurface, which is controlled by an additional parameter, the surface thermal inertia. It is the only parameter to add if we assume that the planet rotates at its equilibrium rotation \citep{Leconte2010}. The planet can, however, be trapped in a spin-orbit resonance as is Mercury. We will address the case of noncircular orbits and nonsynchronized planets in a forthcoming work.

The case of GJ581  is interesting because of its proximity (only 6.2~pc) but also because two planets (e and b) in this system could produce a detectable modulation. Planet `e' is hotter and smaller ($M\sin~i$=1.9~M$_{\oplus}$, $a=0.028$~AU) while planet `b' is cooler but larger ($M\sin i$=16~M$_{\oplus}$, $a=0.04$~AU) and more likely to possess a dense atmosphere. Planet HD40307\,b ($M\sin~i$=4.1~M$_{\oplus}$, $a=0.047$~AU, $M_{*}=0.8M_{\odot}$, $d=13$~pc) represents another fine target. 

\subsection{Transiting planets}
For planets in transit configurations, the radius and inclination are already known. The albedo can be constrained from the spectrum of the planet at the secondary eclipse. However, as is the case of Kepler~10~b \citep{Batalha2011}, the phase curve may be better determined than the secondary eclipse depth due to the short duration of the eclipse. Phase-curve fitting could then provide additional constraints. The phase curve of Kepler~10~b cannot be fitted with thermal emission alone. At the wavelengths observed by Kepler ($0.4-0.9~\mu$m), the reflected light dominates, and the large amplitude of the phase curve points to either a very large Bond albedo ($\sim$0.6) or to a highly anisotropic reflection. 

\subsection{Stellar variability }
\label{variability}
Stellar variability comes from surface brightness inhomogeneities (mainly spots and bright plages), their evolution in time, and their modulation by the stellar differential rotation. These inhomogeneities are produced by magnetic activity and convection, and they produce wavelength-dependent noise and systematic variations that will make the extraction of the planetary phase curve (and thus the constraints on the planet properties less accurate). Using Kepler visible lightcurves of $\sim 150,000$ stars over 33.5~days, \citet{Basri2011} find that more than 70\% of them exhibit relative variations larger than $10^{-3}$. In the mid-infrared, preliminary studies show that this amplitude is expected to be lower by a factor of $\sim 5$ \citep{Bean2010}. To measure the planetary phase with sufficient accuracy, periodic residuals of the stellar variability at the orbital period of the planets have to be lower than the planetary signal (which is larger than $10$~ppm in this work). Period filtering is thus required to lower the amplitude of the noise by at least a factor of $\sim 100$. 

The efficiency of this filtering will depend on the variability power spectrum of the host star, its rotation period compared with the planet orbital period and the level of characterization of this variability that can be achieved. Evaluating the feasibility of this extraction is beyond our expertise and the scope of this paper, but we stress here that it certainly represents the main challenge for this type of high-precision spectro-photometric observations covering two or more orbital periods (days to weeks). It is obviously not an impossible task, at least for some quiet stars. Batalha et al. \citeyearpar{Batalha2011} were for instance able to extract the phase curve of Kepler~10~b, which has an amplitude of less than 10~ppm in the visible range.

\section{Conclusion}

In this study we have modeled the thermal emission of an exoplanet with no atmosphere, on a circular synchronous orbit.  The modulation of this emission with the orbital phase at different wavelengths can be observed for exoplanets in non-transiting configurations, which are known from a previous radial velocity detection. We showed that these infrared multiband phase curves can be used to infer the absence/presence of an atmosphere and, in the airless case, to constrain the radius and albedo of the planet, as well as its orbital inclination. The constraint on the inclination yields a constraint on the mass using the value of $M \sin{i}$ measured by radial velocity. The knowledge of both the radius and the mass can then be used to assess the bulk composition of the planet.
 
Typical precisions of $10^{\circ}$ on the inclination, 0.1 on the albedo, and 10~\% on the radius can be obtained with JWST, and  $15^{\circ}$ on the inclination, 0.2 on the albedo, and 15~\% on the radius with EChO, for nontransiting large terrestrial planets with a period of a few days or less around nearby 10~pc K and M stars. This accuracy is obtained from observing two orbits (for an ideally quiet star) and can be reduced by $\sqrt{N}$ by observing $2~N$ orbits. Assuming that a whole orbit is observed, the accuracy is not very sensitive to the sampling of the orbit. We used ten bands with a 1~$\mu$m width from 5 to 15~$\mu$m and the MIRI's filters, although the number of bands is not critical either. Two bands is the minimum to break the albedo-radius degeneracy. Better results can be obtained for the hottest planets by including shorter wavelengths. 
 
These planets represent an abundant population, as shown by radial velocity surveys and Kepler. More than 90~\% of these hot terrestrial exoplanets do not transit their host star and may only be characterized by their thermal phase curve in the foreseeable future. This method can be applied with EChO or JWST on three already known exoplanets (GJ581~e and b, HD40307~b), and more should be discovered by radial velocity surveys. For each transiting exoplanet with an observable secondary eclipse, there will be statistically ten times more similar objects closer to the Sun and with an observable thermal phase modulation.

The retrieval of $R$, $A$, and $i$ works obviously better for large and hot planets that provide higher planet/star contrast ratios and do so for highly inclined orbits that produce a larger amplitude of the phase variations. Thanks to the distribution function of randomly inclined orbits that is equal to $\sin{i}$, planets are rare whose phase variations are not observable because of a low inclination (only 13~\% with $i<30^{\circ}$).

Unlike secondary eclipse observations, the exposure is not limited by transit duration. Therefore, phase curve fitting can provide better characterize of the planet properties than the eclipse, as in the case of Kepler~10~b. This method implies the ability to achieve spectrophotometry with $10^{-5}$ relative accuracy. This precision must be stable over the duration of the observation used to extract the phase curve. Stellar variability must be characterized and removed in such a way that the residuals do not contaminate the planetary signature. Stability and stellar activity are the most critical issues for phase curve observations. 

Further modeling is required to include the effect of surface roughness and craters on the surface temperature end thermal emission. The case of eccentric planets will be addressed in a forthcoming paper.

\begin{acknowledgements}
A.-S. M. acknowledges fruitful discussions with A. Krones-Martins  about minimization methods and M. Ollivier for its information concerning the instrumental detectors. F. S. acknowledges support from the European Research Council (ERC Grant 209622: E$_3$ARTHs). A.B. acknowledges support from the CNES (Centre National d'Etudes Spatiales).
\end{acknowledgements}

\bibliography{bib.bib}

\begin{thebibliography}{39}
\expandafter\ifx\csname natexlab\endcsname\relax\def\natexlab#1{#1}\fi

\bibitem[{{Ballard} {et~al.}(2010){Ballard}, {Charbonneau}, {Deming},
  {Knutson}, {Christiansen}, {Holman}, {Fabrycky}, {Seager}, \&
  {A'Hearn}}]{Ballard2010}
{Ballard}, S., {Charbonneau}, D., {Deming}, D., {et~al.} 2010, \pasp, 122, 1341

\bibitem[{{Barman} {et~al.}(2005){Barman}, {Hauschildt}, \&
  {Allard}}]{Barman2005}
{Barman}, T.~S., {Hauschildt}, P.~H., \& {Allard}, F. 2005, \apj, 632, 1132

\bibitem[{{Basri} {et~al.}(2011){Basri}, {Walkowicz}, {Batalha}, {Gilliland},
  {Jenkins}, {Borucki}, {Koch}, {Caldwell}, {Dupree}, {Latham}, {Marcy},
  {Meibom}, \& {Brown}}]{Basri2011}
{Basri}, G., {Walkowicz}, L.~M., {Batalha}, N., {et~al.} 2011, \aj, 141, 20

\bibitem[{{Batalha} {et~al.}(2011){Batalha}, {Borucki}, {Bryson}, {Buchhave},
  {Caldwell}, {Christensen-Dalsgaard}, {Ciardi}, {Dunham}, {Fressin},
  {Gautier}, {Gilliland}, {Haas}, {Howell}, {Jenkins}, {Kjeldsen}, {Koch},
  {Latham}, {Lissauer}, {Marcy}, {Rowe}, {Sasselov}, {Seager}, {Steffen},
  {Torres}, {Basri}, {Brown}, {Charbonneau}, {Christiansen}, {Clarke},
  {Cochran}, {Dupree}, {Fabrycky}, {Fischer}, {Ford}, {Fortney}, {Girouard},
  {Holman}, {Johnson}, {Isaacson}, {Klaus}, {Machalek}, {Moorehead},
  {Morehead}, {Ragozzine}, {Tenenbaum}, {Twicken}, {Quinn}, {VanCleve},
  {Walkowicz}, {Welsh}, {Devore}, \& {Gould}}]{Batalha2011}
{Batalha}, N.~M., {Borucki}, W.~J., {Bryson}, S.~T., {et~al.} 2011, \apj, 729,
  27+

\bibitem[{{Bean} {et~al.}(2010){Bean}, {Seifahrt}, {Hartman}, {Nilsson},
  {Wiedemann}, {Reiners}, {Dreizler}, \& {Henry}}]{Bean2010}
{Bean}, J.~L., {Seifahrt}, A., {Hartman}, H., {et~al.} 2010, \apj, 713, 410

\bibitem[{{Belu} {et~al.}(2011){Belu}, {Selsis}, {Morales}, {Ribas}, {Cossou},
  \& {Rauer}}]{Belu2011}
{Belu}, A.~R., {Selsis}, F., {Morales}, J., {et~al.} 2011, Astronomy and
  Astrophysics, 525, A83+

\bibitem[{{Bonfils} {et~al.}(2011){Bonfils}, {Delfosse}, {Udry}, {Forveille},
  {Mayor}, {Perrier}, {Bouchy}, {Gillon}, {Lovis}, {Pepe}, {Queloz}, {Santos},
  {S{\'e}gransan}, \& {Bertaux}}]{Bonfils2011}
{Bonfils}, X., {Delfosse}, X., {Udry}, S., {et~al.} 2011, ArXiv e-prints

\bibitem[{{Bonfils} {et~al.}(2007){Bonfils}, {Mayor}, {Delfosse}, {Forveille},
  {Gillon}, {Perrier}, {Udry}, {Bouchy}, {Lovis}, {Pepe}, {Queloz}, {Santos},
  \& {Bertaux}}]{Bonfils2007}
{Bonfils}, X., {Mayor}, M., {Delfosse}, X., {et~al.} 2007, \aap, 474, 293

\bibitem[{{Burrows} {et~al.}(2010){Burrows}, {Rauscher}, {Spiegel}, \&
  {Menou}}]{Burrows2010}
{Burrows}, A., {Rauscher}, E., {Spiegel}, D.~S., \& {Menou}, K. 2010, \apj,
  719, 341

\bibitem[{{Cash} {et~al.}(2009){Cash}, {Kendrick}, {Noecker}, {Bally},
  {Demarines}, {Green}, {Oakley}, {Shipley}, {Benson}, {Oleson}, {Content},
  {Folta}, {Garrison}, {Gendreau}, {Hartman}, {Howard}, {Hyde}, {Lakins},
  {Leitner}, {Leviton}, {Luquette}, {Oegerley}, {Richon}, {Roberge},
  {Tompkins}, {Tveekrem}, {Woodgate}, {Turnbull}, {Dailey}, {Decker},
  {Dehmohseni}, {Gaugh}, {Glassman}, {Haney}, {Hejal}, {Lillie}, {Lo},
  {O'Conner}, {Oleas}, {Polidan}, {Samuele}, {Shields}, {Shirvanian}, {Soohoo},
  {Tinetti}, {Dorland}, {Dudik}, {Gaume}, \& {Mason}}]{Cash2009}
{Cash}, W., {Kendrick}, S., {Noecker}, C., {et~al.} 2009, in Society of
  Photo-Optical Instrumentation Engineers (SPIE) Conference Series, Vol. 7436,
  Society of Photo-Optical Instrumentation Engineers (SPIE) Conference Series

\bibitem[{{Cockell} {et~al.}(2009){Cockell}, {Herbst}, {L{\'e}ger}, {Absil},
  {Beichman}, {Benz}, {Brack}, {Chazelas}, {Chelli}, {Cottin}, {Coud{\'e} du
  Foresto}, {Danchi}, {Defr{\`e}re}, {den Herder}, {Eiroa}, {Fridlund},
  {Henning}, {Johnston}, {Kaltenegger}, {Labadie}, {Lammer}, {Launhardt},
  {Lawson}, {Lay}, {Liseau}, {Martin}, {Mawet}, {Mourard}, {Moutou}, {Mugnier},
  {Paresce}, {Quirrenbach}, {Rabbia}, {Rottgering}, {Rouan}, {Santos},
  {Selsis}, {Serabyn}, {Westall}, {White}, {Ollivier}, \&
  {Bord{\'e}}}]{Cockell2009}
{Cockell}, C.~S., {Herbst}, T., {L{\'e}ger}, A., {et~al.} 2009, Experimental
  Astronomy, 23, 435

\bibitem[{{Correia} {et~al.}(2010){Correia}, {Couetdic}, {Laskar}, {Bonfils},
  {Mayor}, {Bertaux}, {Bouchy}, {Delfosse}, {Forveille}, {Lovis}, {Pepe},
  {Perrier}, {Queloz}, \& {Udry}}]{Correia2010}
{Correia}, A.~C.~M., {Couetdic}, J., {Laskar}, J., {et~al.} 2010, \aap, 511,
  A21+

\bibitem[{{Cowan} \& {Agol}(2011)}]{Cowan2011}
{Cowan}, N.~B. \& {Agol}, E. 2011, Astrophysical Journal, 726, 82

\bibitem[{{Cowan} {et~al.}(2007){Cowan}, {Agol}, \& {Charbonneau}}]{Cowan2007}
{Cowan}, N.~B., {Agol}, E., \& {Charbonneau}, D. 2007, \mnras, 379, 641

\bibitem[{{Crossfield} {et~al.}(2010){Crossfield}, {Hansen}, {Harrington},
  {Cho}, {Deming}, {Menou}, \& {Seager}}]{Crossfield2010}
{Crossfield}, I.~J.~M., {Hansen}, B.~M.~S., {Harrington}, J., {et~al.} 2010,
  \apj, 723, 1436

\bibitem[{{Deming} {et~al.}(2009){Deming}, {Seager}, {Winn}, {Miller-Ricci},
  {Clampin}, {Lindler}, {Greene}, {Charbonneau}, {Laughlin}, {Ricker},
  {Latham}, \& {Ennico}}]{Deming2009}
{Deming}, D., {Seager}, S., {Winn}, J., {et~al.} 2009, \pasp, 121, 952

\bibitem[{{Fairbairn}(2005)}]{Fairbairn2005}
{Fairbairn}, M.~B. 2005, \jrasc, 99, 92

\bibitem[{{Fischer} {et~al.}(2008){Fischer}, {Marcy}, {Butler}, {Vogt},
  {Laughlin}, {Henry}, {Abouav}, {Peek}, {Wright}, {Johnson}, {McCarthy}, \&
  {Isaacson}}]{Fischer2008}
{Fischer}, D.~A., {Marcy}, G.~W., {Butler}, R.~P., {et~al.} 2008, \apj, 675,
  790

\bibitem[{{Fortney} {et~al.}(2006){Fortney}, {Cooper}, {Showman}, {Marley}, \&
  {Freedman}}]{Fortney2006}
{Fortney}, J.~J., {Cooper}, C.~S., {Showman}, A.~P., {Marley}, M.~S., \&
  {Freedman}, R.~S. 2006, \apj, 652, 746

\bibitem[{{Fortney} {et~al.}(2007){Fortney}, {Marley}, \&
  {Barnes}}]{Fortney2007err}
{Fortney}, J.~J., {Marley}, M.~S., \& {Barnes}, J.~W. 2007, \apj, 668, 1267

\bibitem[{{Gardner} {et~al.}(2006){Gardner}, {Mather}, {Clampin}, {Doyon},
  {Greenhouse}, {Hammel}, {Hutchings}, {Jakobsen}, {Lilly}, {Long}, {Lunine},
  {McCaughrean}, {Mountain}, {Nella}, {Rieke}, {Rieke}, {Rix}, {Smith},
  {Sonneborn}, {Stiavelli}, {Stockman}, {Windhorst}, \& {Wright}}]{Gardner2006}
{Gardner}, J.~P., {Mather}, J.~C., {Clampin}, M., {et~al.} 2006, \ssr, 123, 485

\bibitem[{{Howard} {et~al.}(2011){Howard}, {Marcy}, {Bryson}, {Jenkins},
  {Rowe}, {Batalha}, {Borucki}, {Koch}, {Dunham}, {Gautier}, {Van Cleve},
  {Cochran}, {Latham}, {Lissauer}, {Torres}, {Brown}, {Gilliland}, {Buchhave},
  {Caldwell}, {Christensen-Dalsgaard}, {Ciardi}, {Fressin}, {Haas}, {Howell},
  {Kjeldsen}, {Seager}, {Rogers}, {Sasselov}, {Steffen}, {Basri},
  {Charbonneau}, {Christiansen}, {Clarke}, {Dupree}, {Fabrycky}, {Fischer},
  {Ford}, {Fortney}, {Tarter}, {Girouard}, {Holman}, {Johnson}, {Klaus},
  {Machalek}, {Moorhead}, {Morehead}, {Ragozzine}, {Tenenbaum}, {Twicken},
  {Quinn}, {Isaacson}, {Shporer}, {Lucas}, {Walkowicz}, {Welsh}, {Boss},
  {Devore}, {Gould}, {Smith}, {Morris}, {Prsa}, \& {Morton}}]{Howard2011}
{Howard}, A.~W., {Marcy}, G.~W., {Bryson}, S.~T., {et~al.} 2011, ArXiv e-prints

\bibitem[{{Kane} \& {Gelino}(2011)}]{Kane2011}
{Kane}, S.~R. \& {Gelino}, D.~M. 2011, \apj, 729, 74

\bibitem[{{Knutson} {et~al.}(2009){Knutson}, {Charbonneau}, {Cowan}, {Fortney},
  {Showman}, {Agol}, {Henry}, {Everett}, \& {Allen}}]{Knutson2009}
{Knutson}, H.~A., {Charbonneau}, D., {Cowan}, N.~B., {et~al.} 2009, \apj, 690,
  822

\bibitem[{{Lawson} {et~al.}(2007){Lawson}, {Lay}, {Johnston}, \&
  {Beichman}}]{Lawson2007}
{Lawson}, P.~R., {Lay}, O.~P., {Johnston}, K.~J., \& {Beichman}, C.~A. 2007,
  NASA STI/Recon Technical Report N, 8, 14326

\bibitem[{{Lawson} {et~al.}(2000){Lawson}, {Jakosky}, {Park}, \&
  {Mellon}}]{Lawson2000}
{Lawson}, S.~L., {Jakosky}, B.~M., {Park}, H., \& {Mellon}, M.~T. 2000, \jgr,
  105, 4273

\bibitem[{{Leconte} {et~al.}(2010){Leconte}, {Chabrier}, {Baraffe}, \&
  {Levrard}}]{Leconte2010}
{Leconte}, J., {Chabrier}, G., {Baraffe}, I., \& {Levrard}, B. 2010, \aap, 516,
  A64+

\bibitem[{{L{\'e}ger} {et~al.}(2011){L{\'e}ger}, {Grasset}, {Fegley}, {Codron},
  {Albarede}, {Barge}, {Barnes}, {Cance}, {Carpy}, {Catalano}, {Cavarroc},
  {Demangeon}, {Ferraz-Mello}, {Gabor}, {Griessmeier}, {Leibacher}, {Libourel},
  {Maurin}, {Raymond}, {Rouan}, {Samuel}, {Schaefer}, {Schneider}, {Schuller},
  {Selsis}, \& {Sotin}}]{Leger2011}
{L{\'e}ger}, A., {Grasset}, O., {Fegley}, B., {et~al.} 2011, Icarus, in press

\bibitem[{{Levine} {et~al.}(2009){Levine}, {Lisman}, {Shaklan}, {Kasting},
  {Traub}, {Alexander}, {Angel}, {Blaurock}, {Brown}, {Brown}, {Burrows},
  {Clampin}, {Cohen}, {Content}, {Dewell}, {Dumont}, {Egerman}, {Ferguson},
  {Ford}, {Greene}, {Guyon}, {Hammel}, {Heap}, {Ho}, {Horner}, {Hunyadi},
  {Irish}, {Jackson}, {Kasdin}, {Kissil}, {Krim}, {Kuchner}, {Kwack}, {Lillie},
  {Lin}, {Liu}, {Marchen}, {Marley}, {Meadows}, {Mosier}, {Mouroulis},
  {Noecker}, {Ohl}, {Oppenheimer}, {Pitman}, {Ridgway}, {Sabatke}, {Seager},
  {Shao}, {Smith}, {Soummer}, {Stapelfeldt}, {Tenerell}, {Trauger}, \&
  {Vanderbei}}]{Levine2009}
{Levine}, M., {Lisman}, D., {Shaklan}, S., {et~al.} 2009, ArXiv e-prints

\bibitem[{{Mayor} {et~al.}(2009{\natexlab{a}}){Mayor}, {Bonfils}, {Forveille},
  {Delfosse}, {Udry}, {Bertaux}, {Beust}, {Bouchy}, {Lovis}, {Pepe}, {Perrier},
  {Queloz}, \& {Santos}}]{Mayor2009a}
{Mayor}, M., {Bonfils}, X., {Forveille}, T., {et~al.} 2009{\natexlab{a}}, \aap,
  507, 487

\bibitem[{{Mayor} {et~al.}(2011){Mayor}, {Marmier}, {Lovis}, {Udry},
  {S{\'e}gransan}, {Pepe}, {Benz}, {Bertaux}, {Bouchy}, {Dumusque}, {Lo Curto},
  {Mordasini}, {Queloz}, \& {Santos}}]{Mayor2011}
{Mayor}, M., {Marmier}, M., {Lovis}, C., {et~al.} 2011, ArXiv e-prints

\bibitem[{{Mayor} {et~al.}(2009{\natexlab{b}}){Mayor}, {Udry}, {Lovis}, {Pepe},
  {Queloz}, {Benz}, {Bertaux}, {Bouchy}, {Mordasini}, \&
  {Segransan}}]{Mayor2009b}
{Mayor}, M., {Udry}, S., {Lovis}, C., {et~al.} 2009{\natexlab{b}}, \aap, 493,
  639

\bibitem[{{Rivera} {et~al.}(2010){Rivera}, {Laughlin}, {Butler}, {Vogt},
  {Haghighipour}, \& {Meschiari}}]{Rivera2010}
{Rivera}, E.~J., {Laughlin}, G., {Butler}, R.~P., {et~al.} 2010, \apj, 719, 890

\bibitem[{{Seager} \& {Deming}(2009)}]{Seager2009}
{Seager}, S. \& {Deming}, D. 2009, \apj, 703, 1884

\bibitem[{{Selsis} {et~al.}(2011){Selsis}, {Wordsworth}, \&
  {Forget}}]{Selsis2011}
{Selsis}, F., {Wordsworth}, R.~D., \& {Forget}, F. 2011, \aap, 532, A1+

\bibitem[{{Showman} {et~al.}(2009){Showman}, {Fortney}, {Lian}, {Marley},
  {Freedman}, {Knutson}, \& {Charbonneau}}]{Showman2009}
{Showman}, A.~P., {Fortney}, J.~J., {Lian}, Y., {et~al.} 2009, \apj, 699, 564

\bibitem[{{Tinetti} {et~al.}(2011){Tinetti}, {Beaulieu}, {Henning}, {Meyer},
  {Micela}, {Ribas}, {Stam}, {Swain}, {Krause}, {Ollivier}, {Pace}, {Swinyard},
  {Aylward}, {van Boekel}, {Coradini}, {Encrenaz}, {Snellen},
  {Zapatero-Osorio}, {Bouwman}, {Y-K.~Cho}, {Coud{\'e} du Foresto}, {Guillot},
  {Lopez-Morales}, {Mueller-Wodarg}, {Palle}, {Selsis}, {Sozzetti}, {Ade},
  {Achilleos}, {Adriani}, {Agnor}, {Afonso}, {Allende Prieto}, {Bakos},
  {Barber}, {Barlow}, {Bernath}, {Bezard}, {Bord{\'e}}, {Brown}, {Cassan},
  {Cavarroc}, {Ciaravella}, {Cockell}, {Coustenis}, {Danielski}, {Decin}, {De
  Kok}, {Demangeon}, {Deroo}, {Doel}, {Drossart}, {Fletcher}, {Focardi},
  {Forget}, {Fossey}, {Fouqu{\'e}}, {Frith}, {Galand}, {Gaulme}, {Gonz{\'a}lez
  Hern{\'a}ndez}, {Grasset}, {Grassi}, {Grenfell}, {Griffin}, {Griffith},
  {Gr{\"o}zinger}, {Guedel}, {Guio}, {Hainaut}, {Hargreaves}, {Hauschildt},
  {Heng}, {Heyrovsky}, {Hueso}, {Irwin}, {Kaltenegger}, {Kervella}, {Kipping},
  {Koskinen}, {Kov{\'a}cs}, {La Barbera}, {Lammer}, {Lellouch}, {Leto}, {Lopez
  Morales}, {Lopez Valverde}, {Lopez-Puertas}, {Lovis}, {Maggio}, {Maillard},
  {Maldonado Prado}, {Marquette}, {Martin-Torres}, {Maxted}, {Miller},
  {Molinari}, {Montes}, {Moro-Martin}, {Moses}, {Mousis}, {Nguyen Tuong},
  {Nelson}, {Orton}, {Pantin}, {Pascale}, {Pezzuto}, {Pinfield}, {Poretti},
  {Prinja}, {Prisinzano}, {Rees}, {Reiners}, {Samuel}, {Sanchez-Lavega}, {Sanz
  Forcada}, {Sasselov}, {Savini}, {Sicardy}, {Smith}, {Stixrude}, {Strazzulla},
  {Tennyson}, {Tessenyi}, {Vasisht}, {Vinatier}, {Viti}, {Waldmann}, {White},
  {Widemann}, {Wordsworth}, {Yelle}, {Yung}, \& {Yurchenko}}]{Tinetti2011}
{Tinetti}, G., {Beaulieu}, J.~P., {Henning}, T., {et~al.} 2011, ArXiv e-prints

\bibitem[{{Vogt} {et~al.}(2010){Vogt}, {Wittenmyer}, {Butler}, {O'Toole},
  {Henry}, {Rivera}, {Meschiari}, {Laughlin}, {Tinney}, {Jones}, {Bailey},
  {Carter}, \& {Batygin}}]{Vogt2010}
{Vogt}, S.~S., {Wittenmyer}, R.~A., {Butler}, R.~P., {et~al.} 2010, \apj, 708,
  1366

\bibitem[{{Wright} {et~al.}(2004){Wright}, {Rieke}, {Colina}, {van Dishoeck},
  {Goodson}, {Greene}, {Lagage}, {Karnik}, {Lambros}, {Lemke}, {Meixner},
  {Norgaard}, {Oloffson}, {Ray}, {Ressler}, {Waelkens}, {Wright}, \&
  {Zhender}}]{Wright2004}
{Wright}, G.~S., {Rieke}, G.~H., {Colina}, L., {et~al.} 2004, in Society of
  Photo-Optical Instrumentation Engineers (SPIE) Conference Series, Vol. 5487,
  Society of Photo-Optical Instrumentation Engineers (SPIE) Conference Series,
  ed. {J.~C.~Mather}, 653--663

\end{thebibliography}
\bibliographystyle{aa}


\clearpage

\clearpage

\clearpage

\clearpage

\begin{table*}
\centering 
\begin{tabular}{| c | c | c | c | c | c | c | c | c |} 
\hline

\multicolumn{2}{|c|}{} & \multicolumn{3}{|c|}{5-11 $\mu$m} & \multicolumn{3}{|c|}{11-15 $\mu$m}\\
\hline
Telescope & Type of detectors & RON & dark current & full-well capacity & RON & dark current & full-well capacity\\
\hline
EChO & Si:As & 12 & 0.1 & 2$\cdot 10^5$ & 19  & 0.1 & 2$\cdot 10^5$\\
\hline
EChO & MCT+Si:As & 1000 & 500 & 37$\cdot 10^6$ & 12 & 0.1 & 2$\cdot 10^5$\\
\hline

\end{tabular}
\caption{Characteristics of detectors used for EChO. The readout noise RMS is in e$^-$/pixel, the dark current in e$^-$/s/pixel, and the full-well capacity in e$^-$. Only all-Si:As are used in the results shown in this paper, MCT+Si:As giving comparable results.} 
\label{detectors-EChO} 
\end{table*}
\clearpage

\begin{table}[p]
\centering 
\begin{tabular}{| c | c | c | c |} 
\hline\hline
\backslashbox{$a$ (AU)}{$M$ ($M_{\odot}$)} & 0.2 & 0.5 & 0.8 \\ 
\hline 
\multirow{2}{*}{0.02} & T=596~K& 983~K& 1641~K \\ 
	& P=2.31~d & 1.46~d & 1.15~d\\
\hline
\multirow{2}{*}{0.04} & T=421~K & 695~K & 1160~K \\ 
	& P=6.53~d & 4.13~d & 3.26~d\\
\hline 
\end{tabular}
\caption{Equilibrium temperatures (K) and orbital periods (days) for different stellar masses and orbital distances. Temperatures are calculated with a Bond albedo of 0.1.} 
\label{cases} 
\end{table}

%
%
%
%
%

\clearpage

\clearpage



\clearpage

\clearpage

\clearpage

\clearpage

\end{document}